\documentclass[prodmode,acmtist]{acmsmall} 

\usepackage{amsmath}

\usepackage{algorithm}
\usepackage[noend]{algpseudocode}
\algloopdefx{ParFor}{\textbf{par-for} }   
\algrenewcommand\algorithmicindent{1.0em} 

\newcommand{\V}[1]{\boldsymbol{#1}}
\newcommand{\M}[1]{#1}

\acmVolume{9}
\acmNumber{4}
\acmArticle{39}
\acmYear{2014}
\acmMonth{12}

\begin{document}

\markboth{Y. Wang et al.}{Peacock: Learning Long-Tail Topic Features for Industrial Applications}
\title{Peacock: Learning Long-Tail Topic Features for Industrial Applications}
\author{YI WANG
\affil{Tencent} XUEMIN ZHAO \affil{Tencent} ZHENLONG SUN
\affil{Tencent} HAO YAN \affil{Tencent} LIFENG WANG \affil{Tencent}
ZHIHUI JIN \affil{Tencent} LIUBIN WANG\affil{Tencent} YANG
GAO\affil{School of Computer Science and Technology, Soochow
University} CHING LAW\affil{Tencent} JIA ZENG\affil{School of
Computer Science and Technology, Soochow University \& Huawei Noah's
Ark Lab}}

\begin{abstract}
Latent Dirichlet allocation (LDA) is a popular topic modeling technique in academia but less so in industry,
especially in large-scale applications involving search engine and online advertising systems.
A main underlying reason is that the topic models used have been too small in scale to be useful;
for example,
some of the largest LDA models reported in literature have up to $10^3$ topics,
which cover difficultly the long-tail semantic word sets.
In this paper,
we show that the number of topics is a key factor that can significantly boost the utility of topic-modeling systems.
In particular,
we show that a ``big" LDA model with at least $10^5$ topics inferred from $10^9$ search queries can achieve a
significant improvement on industrial search engine and online advertising systems,
both of which serving hundreds of millions of users.
We develop a novel distributed system called Peacock to learn big LDA models from big data.
The main features of Peacock include hierarchical distributed architecture,
real-time prediction and topic de-duplication.
We empirically demonstrate that the Peacock system is capable of providing significant benefits via highly
scalable LDA topic models for several industrial applications.
\end{abstract}







\begin{bottomstuff}
Author's addresses:
Y. Wang, X. Zhao, Z. Sun, H. Yan, Z. Jin, L. Wang and C. Law, Tencent;
Y. Gao and J. Zeng, School of Computer Science and Technology, Soochow University, Shuzhou 215006, China \&
Collaborative Innovation Center of Novel Software Technology and Industrialization;
J. Zeng, Huawei Noah's Ark Lab, Hong Kong. J. Zeng is the corresponding author: zeng.jia@acm.org.
\end{bottomstuff}

\maketitle

\section{Introduction} \label{sec:intro}

In academia,
latent Dirichlet allocation (LDA)~\cite{blei} is a popular topic modeling technique,
which is an unsupervised learning algorithm to infer semantic word sets called topics.
However,
very few successes of LDA have been reported in industry.
The major reason is that the largest LDA models reported in literature~\cite{gpu-lda,ad-lda,as-lda,plda,plda+,mrlda} have up to $10^3$ topics,
which cannot cover completely the long-tail semantic word sets in big data.
Industrial applications like search engine and online advertising require the capability of learning many semantic word sets (or topics)
that cover a large part of human knowledge, in particular, the long-tail part.
As reported by Linguistic Data Consortium (LDC),
there are millions of vocabulary words in either English, Chinese, Spanish or Arabic~\cite{gigaword}.
Taking polysemy and synonyms into consideration,
a rough estimate of the number of word senses is close to the same magnitude of vocabulary words, i.e.,
$10^5$ or $10^6$ topics for semantics of long-tail word sets.

To the best of our knowledge,
the number of topics in applications mentioned above is around two to three orders of magnitude larger than
that in the current state-of-the-art~\cite{smola,Ahmed:12}.
The effectiveness of the large number of topics is inspired by~\cite{pfp},
which proposes a MapReduce-based \emph{frequent itemset mining} algorithm to find the long-tail word sets.
We notice that there is a word set containing the words ``whorf piraha chomsky anthropology linguistics''.
Using web search,
we find that this word set has a clear semantic meaning on the research by Whorf and Chomsky,
which is related to anthropology and linguistics based on their study of the Piraha language.
However,
it is a regret that the \emph{frequent itemset mining} algorithm cannot interpret new documents out of the training corpus.
Fortunately,
LDA overcomes this shortcoming and can infer highly interpretable and semantically coherent topics from new documents.

\begin{figure}
\centerline{\includegraphics[width=0.8\linewidth]{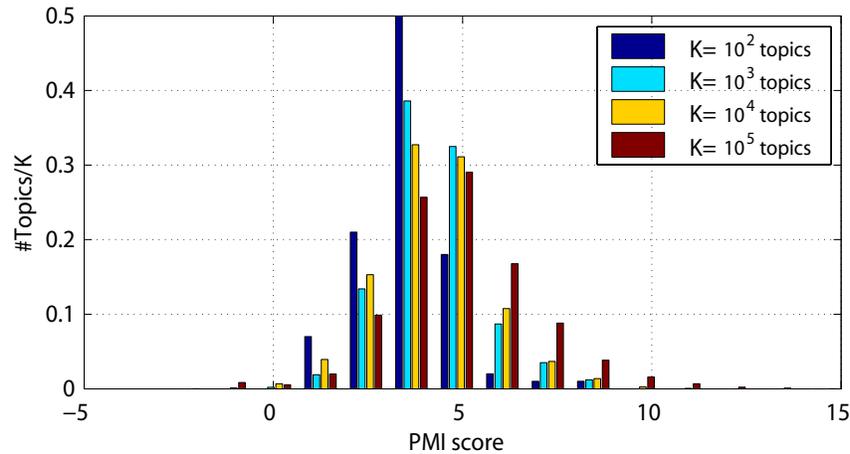}}
\caption{The percentage of topics over the PMI score by LDA models with different number of topics $K$.}
\label{fig:pmi}
\end{figure}

To illustrate the advantage of the large number of topics $K$,
we use the point-wise mutual information (PMI) to measure the interpretability or semantic coherence of topics~\cite{pmi}.
The higher PMI score corresponds to a better topic quality.
First,
we learn LDA models with different number of topics $K \in \{10^2, 10^3, 10^4, 10^5\}$
on the SOSO data set described in Subsection~\ref{sec:data}.
Second,
we calculate the PMI score of each topic and obtain the topic histograms over the PMI score bucket of different LDA models.
Figure~\ref{fig:pmi} shows the percentage of topics over the PMI score
by four LDA models with $10^2$, $10^3$, $10^4$ and $10^5$ topics, respectively.
With the increase of topics from $10^2$ to $10^5$,
we see that the topic histograms shift toward the larger mean PMI score,
i.e.,
more and more topics have higher PMI scores.
This phenomenon suggests that larger LDA models tend to encode more interpretable and semantically coherent topics.

In this paper,
we confirm that big LDA models with at least $10^5$ topics can achieve a significant improvement in two industrial
applications such as search engine and online advertising systems.
This finding motivates us to pursue scalable topic modeling systems for big data.
To achieve this goal,
we develop a hierarchical distributed learning system called Peacock that can generate a much larger number of topics than before.
For example,
Peacock can learn at least $10^5$ topics from $10^9$ search queries.
This improvement is nontrivial and raises many new technical challenges.
First,
how to make the system scalable to process big query data as well as LDA parameters with fault tolerance?
Second,
how to do real-time topic prediction for new queries and how to remove duplicate topics to obtain high-quality ones?
Finally,
how to integrate big LDA models into existing search engine and online advertising systems for a better performance?
We address these technical issues and summarize our contributions as follows:
\begin{itemize}
\item
We design a new hierarchical distributed architecture including
model parallelism to handle a large number of LDA parameters as well as data parallelism to handle massive training corpora.
We also use the pipeline and lock-free techniques to reduce communication and synchronization costs.
This architecture runs on a computer cluster including thousands of CPU cores,
which can learn $\ge 10^5$ topics from $\ge 10^9$ search queries,
around two orders of magnitude larger than the current state-of-the-art reported in literature~\cite{smola,Ahmed:12}.
\item
When performing topic modeling for big data,
we solve two new practical problems in real-world applications: real-time prediction and topic de-duplication.
A new real-time prediction algorithm called RT-LDA is developed to infer the topic distributions of unseen queries in
search engine and online advertising systems.
As far as topic de-duplication is concerned,
we use two methods:
1) Learning asymmetric Dirichlet priors~\cite{rethinking};
2) Clustering similar topics by their $L_1$-similarities.
\item
We examine the effectiveness of big LDA models in two online industrial applications:
search engine and online advertising systems.
The performance improvements in both systems grow with the increasing number of the learned topics.
We observe a significant improvement on search relevance when the number of topics increases from $10^3$ to $10^4$.
Also,
the topic features significantly improve the accuracy of ad click-through rate prediction when the number of topics increases from $10^4$ to $10^5$.
\end{itemize}

The rest of this paper is organized as follows.
In the next section,
we discuss related work.
In Section~\ref{sec:peacock},
we present the hierarchical distributed architecture,
real-time prediction and topic de-duplication in the proposed Peacock system.
In Section~\ref{sec:parallel},
we show that Peacock is more scalable to the larger number of topics than the state-of-the-art industrial solution
Yahoo!LDA~\cite{smola,Ahmed:12}.
Section~\ref{sec:industry} shows two online industrial applications of Peacock:
search engine and online advertising systems.
In Section~\ref{sec:conclusion},
we make conclusions and envision future work.

\section{Related Work and Discussion}

There are five categories of batch inference algorithms proposed to estimate LDA parameters:
variational Bayes (VB)~\cite{blei},
Gibbs sampling (GS)~\cite{gs,fastlda,sparselda},
expectation propagation (EP)~\cite{ep},
belief propagation (BP)~\cite{Zeng:11,Zeng:12},
and collapsed variational Bayes (CVB or CVB0 with zero-order approximation)~\cite{Teh:06,Asuncion:09,Sato:12}.
Except for GS,
all other inference methods are based on the coordinate ascent algorithm~\cite{Freitas:01,Murphy:book},
which first calculates the topic posterior distribution over each word token,
and then updates the parameters based on the inferred posterior distribution.
Although CVB0 and BP converge much faster and produce higher held-out log-likelihood~\cite{Wallach:09} than GS,
they require storing the posterior probability matrix of all words in memory.
The size of this matrix increases linearly with the number of unique document/word pairs and the number of topics.
It is difficult to distribute this big matrix to a common computer cluster when the number of words and the number of topics are very large.
In addition,
CVB0 and BP store the parameters of document-topic and topic-word distributions in double precision,
consuming more memory to handle sparse data sets.
Most parallel inference solutions of LDA choose batch GS algorithms
because they are more memory-efficient than other algorithms~\cite{smola,Ahmed:12,fastlda,plda,ad-lda,plda+,gpu-lda}.
For example,
GS does not need to maintain the large posterior matrix in memory.
In addition,
GS stores LDA parameters using the integer type by sparse matrices,
and often obtains higher topic modeling accuracy (e.g., higher held-out log-likelihood) than VB~\cite{gs}.
So,
GS is generally agreed to be a more scalable choice in many parallel LDA solutions.
We will discuss how to distribute an accelerated GS algorithm with low time and space complexities called SparseLDA in Subsection~\ref{sec:sparselda}.

Previous distributed LDA systems have explored two main architectures:
1) parallel computing using processors tightly coupled with shared memory~\cite{gpu-lda},
and 2) distributed computing using processors loosely connected via network~\cite{ad-lda,as-lda,plda,plda+,mrlda,Yan:13,Yan:14}.
Yahoo!LDA~\cite{smola,Ahmed:12} can be viewed as a hybrid parallel architecture
using the shared memory technique over multiple machines based on the \emph{memcached} technique.
Mr. LDA~\cite{mrlda} distributes the batch VB algorithm to the MapReduce framework,
which requires frequent I/O operations causing the slow speed.
However,
all reported distributed LDA systems learn up to $10^3$ topics,
while $10^3$ might be far less than the real number of semantics or word senses in human language.
On the other hand,
LDA is a non-negative matrix factorization method~\cite{Buntine:05,Zeng:11},
so that many parallel matrix factorization architectures can be used.
However,
recent parallel matrix factorization architectures~\cite{Gemulla:11,Zhuang:13} have difficulty in learning $10^5$ topics
because they focus on only low-rank approximation to the big sparse matrix. The rank is often very low such as $10^2 \sim 10^3$,
where the rank in matrix factorization has the same meaning
with the number of topics in LDA.
Unlike previous distributed LDA solutions,
Peacock introduces both data and model parallelism for big data ($\ge 10^9$ documents)
and big LDA models ($\ge 10^9$ parameters including topic assignment vector and topic-word matrix) within a hierarchical distributed architecture,
which uses pipeline and lock-free techniques to reduce both communication and synchronization costs.
In addition,
Peacock has two new components,
real-time prediction and topic de-duplication,
which play important roles in real-world applications.

Recently,
online LDA algorithms~\cite{Hoffman:10,online-bp,Mimno:12,Patterson:13,Broderick:13,Foulds:13}
have attracted intensive research interests with two reasons.
First,
except for online Bayesian updating~\cite{Broderick:13},
most online LDA algorithms combine the stochastic optimization framework~\cite{Robbins:51} with the corresponding batch LDA algorithms,
which theoretically can converge to the local optimal point of the LDA objective function.
Second,
online algorithms partition the entire data set into several mini-batches.
They load each mini-batch in memory for online processing,
and discard it after one look.
This streaming method significantly reduces the memory consumption for big data and big models.
However,
given the same amount of training samples,
batch algorithms converge significantly faster and yield higher held-out log-likelihood than online counterparts~\cite{Zeng:13,online-bp}.
The main reason is that the convergence rate of stochastic algorithms is slower
than that of batch algorithms which has been discussed in~\cite{Schmidt:13}.
If we have enough computing resources,
it is better to distribute batch algorithms than online ones.
If the memory is limited and the data come in the streaming manner,
we prefer distributing online algorithms.
For example,
D-SGLD~\cite{Ahn:14} is based on a stochastic gradient MCMC (Monte Carlo Markov Chain) and is distributed with adaptive
load balance by making the faster workers work longer until the slower workers finish their tasks.
POBP~\cite{Yan:13} parallelizes the online belief propagation algorithm,
and has a dynamic communication scheduling scheme to reduce the overall cost in processing big data streams.
As discussed above,
both batch and online algorithms have their advantages and disadvantages.
Although Peacock is designed for distributed batch algorithms,
its architecture can be readily extended to distributed online algorithms for streaming data.
In this case,
Peacock just replaces batch algorithms with online counterparts without changing the hierarchical architecture,
which will be studied in our future work.

Other topic models such as hierarchical Dirichlet processes (HDP)~\cite{hdp} and
author-topic models (ATM)~\cite{Steyvers:04} have similar batch inference algorithms with LDA
such as GS and VB~\cite{hdp,Hoffman:13}.
These inference algorithms estimate similar parameters as those in LDA,
for example,
the multinomial parameters for topic-word distributions.
Therefore,
the parallel implementation of these inference algorithms can be also deployed in Peacock.
For instance,
ATM can be implemented in Peacock by replacing the multinomial parameters over documents with those over authors~\cite{Zeng:11}.

\section{The Peacock System} \label{sec:peacock}

\begin{table}[t]
\centering
\tbl{Definitions of Notation.}
{\begin{tabular}{|l|l|}
\hline
$1\leq d\leq D$ & Document index \\ \hline
$1\leq v\leq V$ & Vocabulary word index \\ \hline
$1\leq k\leq K$ & Topic index \\ \hline
$1\leq m\leq M$ & Model and data shard index \\ \hline
$1\leq c\leq C$ & Configuration index \\ \hline
$x_{ivd}$       & Word token in bag-of-word representation \\ \hline
$z_{ivd}$       & Topic label for each word token \\ \hline
$\neg ivd$      & All word tokens except the $ivd$ word token \\ \hline
$\Theta_{dk}$   & Number of tokens in document $d$ assigned to the topic $k$ \\ \hline
$\theta_{dk}$   & Multinomial parameters for document-topic distribution \\ \hline
$\Phi_{vk}$     & Number of tokens of type $v$ assigned to the topic $k$ \\ \hline
$\phi_{vk}$     & Multinomial parameters for topic-word distribution \\ \hline
$\Psi_{k}$      & $\sum_{v}{\Phi_{vk}}$ \\ \hline
$\alpha_k$      & Asymmetric Dirichlet hyperparameter \\ \hline
$\beta$         & Symmetric Dirichlet hyperparameter \\ \hline
$l_d$           & Document length for estimating asymmetric hyperparameter $\alpha_k$ \\ \hline
$\Omega_{kn}$   & Count matrix for estimating asymmetric hyperparameter $\alpha_k$ \\ \hline
$T$             & Number of slots in the pipeline communication \\ \hline
$L$             & Packet size in the pipeline communication \\ \hline
\end{tabular}}
\label{notations}
\end{table}

In this section,
we first introduce an accelerated GS algorithm called SparseLDA~\cite{sparselda} for learning LDA.
Then,
we show how to distribute the GS algorithm in the hierarchical architecture with model and data parallelism
to handle the large number of LDA parameters and training samples.
In this new architecture,
we focus on implementing the following techniques:
1) the distributed GS algorithm,
2) pipeline efficient communication,
3) lock-free synchronization,
and 4) fault tolerance.
Finally,
we introduce how to solve two new problems of big LDA models in real-world industrial applications:
1) real-time topic prediction of new unseen queries/documents,
and 2) topic de-duplication for better quality and performance.
Table~\ref{notations} summarizes some important notations used in this paper.

\subsection{Accelerated Gibbs Sampling (GS) Inference for LDA} \label{sec:sparselda}

LDA allocates a set of thematic topic labels,
$\V{z} = \{z_{ivd}=k\}$,
to explain the word tokens,
$\V{x}_{D \times V} = \{x_{ivd} = \{0,1\} \}$,
in the document-word co-occurrence matrix $\V{x}_{D \times V}$,
where $i$ is the word token index,
$1 \le v \le V$ denotes the word index in the vocabulary,
$1 \le d \le D$ denotes the document index in the corpus,
and $1 \le k \le K$ denotes the topic index.
Usually,
the number of topics $K$ is provided by users.
The objective of LDA is to maximize the joint probability $P(\V{x},\theta,\phi|\alpha,\beta)$,
where $\theta_{D \times K}$ and $\phi_{V \times K}$
are two non-negative matrices of multinomial parameters for document-topic and topic-word distributions,
satisfying $\sum_k \theta_{dk} = 1$ and $\sum_v \phi_{vk} = 1$.
Both multinomial matrices are generated by two Dirichlet distributions with hyperparameters $\alpha$ and $\beta$.

Since we aim to learn $K \ge 10^5$ topic from $D \ge 10^9$ search queries,
we choose to distribute an accelerated sparse Gibbs sampling (GS) inference algorithm called SparseLDA~\cite{sparselda},
whose time and space complexities are insensitive to the number of topics $K$.
In GS,
the memory is used to maintain three LDA parameter count matrices:
a matrix $\Phi_{V \times K}$ in which each element is the total number of the vocabulary word $v$ assigned to the topic $\V{z}_{ivd} = k$,
a matrix $\Theta_{D \times K}$ in which each element is the total number of topic $\V{z}_{ivd} = k$ assignments in each document $d$,
and a count vector $\Psi_k = \sum_{v=1}^{V} \Phi_{vk}$,
in which each element is the number of topic $k$ assignments in the training corpus.
The relation between LDA multinomial parameters and count matrices are as follows:
\begin{gather}
\label{theta}
\theta_{dk} = \frac{\Theta_{dk} + \alpha}{\sum_{k=1}^K \Theta_{dk} + K\alpha}, \\
\label{phi}
\phi_{vk} = \frac{\Phi_{vk} + \beta}{\Psi_k + V\beta}.
\end{gather}
In each iteration,
the GS algorithm updates the topic assignment $z_{ivd}=k$ of every observed word token
$x_{ivd}=1$ in the training corpus by randomly drawing a topic $z_{ivd}=k$ from the
collapsed posterior distribution,
\begin{align} \label{eq:gibbs-sampling}
P(z_{ivd}=k, x_{ivd}=1 \mid \V{z}_{\neg ivd}, \V{x}_{\neg ivd},
\alpha, \beta) \propto  \frac{\Phi_{vk}^{\neg ivd} +
\beta}{\Psi_k^{\neg ivd} + V\beta} \left(\Theta_{dk}^{\neg ivd} +
\alpha \right),
\end{align}
where $\neg ivd$ means that the corresponding word token and topic $z_{ivd}=k$ is excluded from the count matrices.
After the new topic assignment $z_{ivd} = k$ is sampled,
the corresponding elements in the count matrices $\Psi_k$, $\Phi_{vk}$ and $\Theta_{dk}$ are updated immediately.
SparseLDA divides Equation~\eqref{eq:gibbs-sampling} into three parts:
\begin{align} \label{eq:sparseLDA}
P(z_{ivd}=k, x_{ivd}=1 \mid \V{z}_{\neg ivd}, \V{x}_{\neg ivd},
\alpha, \beta) \propto \frac{\alpha\beta}{\Psi_{k}^{\neg
ivd}+V\beta}+\frac{\Theta_{dk}^{\neg ivd}\beta}
{\Psi_{k}^{\neg
ivd}+V\beta} \notag \\
+\frac{(\alpha+\Theta_{dk}^{\neg ivd})\Phi_{vk}^{\neg ivd}}{\Psi_{k}^{\neg
ivd}+V\beta}.
\end{align}
Due to sparsity of the topic posterior probability $P(z_{ivd}=k, x_{ivd}=1 \mid \V{z}_{\neg ivd}, \V{x}_{\neg ivd}, \alpha, \beta)$,
randomly sampling these three parts does not need to calculate $K$ times.
As a result,
SparseLDA has a time complexity insensitive to the number of topics $K$.
Moreover,
it has a low space complexity because
it stores only the topic assignment vector $\V{z}$ rather than the matrix $\Theta_{D \times K}$ in memory,
where the size of $\V{z}$ is equal to the total number of word tokens in corpus,
which is irrelevant with the number of topics $K$.
SparseLDA re-organizes the corresponding $\V{z}_{ivd}$ into the document-specific vector $\Theta_{dk}$ on the fly.
If a vocabulary word $v$ has a total of $N \ll K$ occurrences in training corpus,
the $v$th row of the parameter matrix $\Phi_{vk}$ only needs to store up to $N$ rather than $K$ values.

\subsection{Hierarchical Distributed Architecture} \label{sec:hierarchical}

\begin{figure}
\centerline{\includegraphics[width=0.5\linewidth]{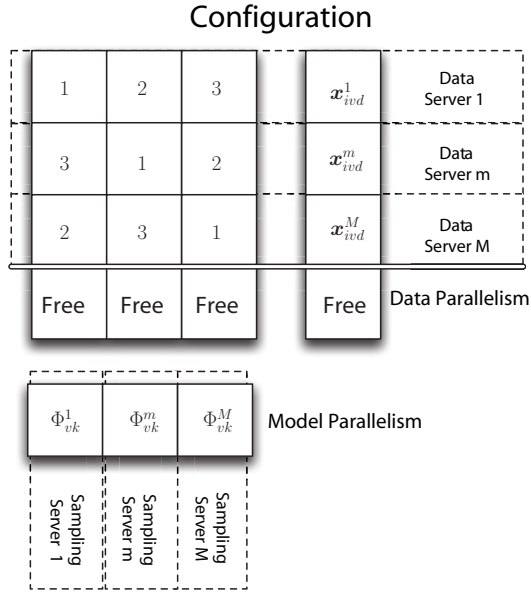}}
\caption{One configuration in the first layer of the hierarchical distributed architecture.}
\label{fig:config}
\end{figure}

\begin{figure}
\centerline{\includegraphics[width=1.0\linewidth]{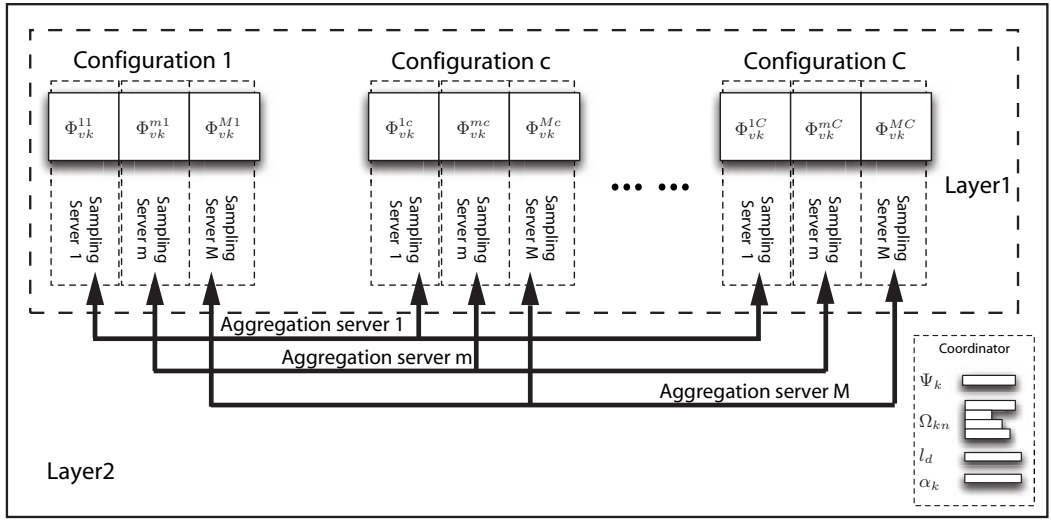}}
\caption{The hierarchical distributed architecture in Peacock.
The first layer contains multiple configurations in Figure~\ref{fig:config}.
The second layer contains $M$ aggregation servers and
one coordinator server for global parameter synchronization and asymmetric prior estimation.
This architecture can readily scale up to hundreds of machines having thousands of cores to learn at least $10^5$ topics from $10^9$ search queries.}
\label{fig:layer2}
\end{figure}

The key challenge is to store the word tokens $\V{x}$,
the topic assignment vector $\V{z}$,
and the large LDA parameter matrix $\Phi_{V \times K}$,
when $V \ge 10^5$, $K \ge 10^5$ and $D \ge 10^9$ in industrial applications.
For example,
the count matrix $\Phi_{V \times K}$ alone takes at least tens of gigabytes when learning $10^5$ topics,
while modern computer clusters are composed of commodity computers~\cite{mapreduce} with few gigabytes of memory (e.g., $2$GB memory).
Therefore,
we propose the hierarchical distributed architecture to solve this large-scale problem,
which contains a configuration of servers to handle both big data and big LDA model in Figure~\ref{fig:config}.
More specifically,
the distributed GS algorithm can be executed by a configuration of the following servers:
\begin{enumerate}
\item
{\em Model parallelism}:
We partition the parameter $\Phi_{V \times K}$ matrix by rows $1 \le v \le V$ into $1 \le m \le M$ model shards,
$\{\Phi_{vk}^{1}, \ldots, \Phi_{vk}^m, \ldots, \Phi_{vk}^M\}$.
We use $M$ {\em sampling servers},
where the $m$th sampling server maintains the $\Phi_{vk}^m$ shard and local copies of $\Psi_k^{m}$ and $\alpha^m_k$ in memory.
The value of $M$ should make each model shard $\Phi_{vk}^m$ small enough to fit in the memory of a sampling server.
Each sampling server runs the SparseLDA algorithm to update the topic assignments $\V{z}^m_{ivd}$
in training blocks sent from the data servers.
The sampling algorithm will update the $\Phi_{vk}^{m}$ shard at $m$th sampling server. At the end of the training process,
all sampling servers output their $\Phi_{vk}^{m}$ model shards.
\item
{\em Data parallelism}:
We partition the word tokens $\V{x}_{D \times V}$ and their topic assignments $\V{z}$ by rows into $1 \le m \le M$ shards,
$\{\V{x}_{idv}^1, \V{z}_{idv}^1, \ldots, \V{x}_{idv}^m,
\V{z}_{idv}^m, \ldots, \V{x}_{idv}^M, \V{z}_{idv}^M\}$.
We use $M$ {\em data servers},
where each loads a data shard $\V{x}_{idv}^m$ and the corresponding $\V{z}_{idv}^m$ shard in memory.
The value of $M$ should make each data shard and its corresponding $\V{z}$ shard
small enough to fit in the memory of a data server.
The data servers send word tokens and their topic assignments shards to the corresponding sampling servers,
which update the topic assignments $\V{z}_{idv}^m$ and the model parameters $\Phi_{vk}^m$.
After processing each segment of data shards,
the sampling servers send back the changed topic assignments $\Delta\V{z}_{idv}^m$ to data servers,
which write $\Delta\V{z}_{idv}^m$ back to disks for fault recovery.
\end{enumerate}
As a result,
the entire data has been partitioned into $M \times M$ blocks and each data server stores $M$ blocks.
If the sampling server simultaneously obtains the same column of data segment sent by all data servers,
it will read and write the topic assignments of the same vocabulary word under race conditions.
For example in Figure~\ref{fig:config},
if the sampling server $1$ receives simultaneously three data blocks $\{1,3,2\}$ sent by three data servers,
it has a higher likelihood to change the topic assignments of the same vocabulary words,
which causes serious read/write locks and I/O delays.
To solve this problem,
we design a lock-free parallel strategy similar to~\cite{gpu-lda,Gemulla:11,Zhuang:13}.
The sampling servers process blocks on the main diagonal sent by the data servers.
Since only $\Phi_{vk}^{1}$ and $\{\V{x}_{ivd}^{1}, \V{z}_{ivd}^1\}$ are required for processing the first data block,
$\Phi_{vk}^{2}$ and $\{\V{x}_{ivd}^{2}, \V{z}_{ivd}^2\}$ for the second, and
$\Phi_{vk}^{3}$ and $\{\V{x}_{ivd}^{3}, \V{z}_{ivd}^3\}$ for the third,
these three blocks can be processed simultaneously without conflicts in accessing $\Phi_{vk}$ and $\{\V{x}_{ivd}, \V{z}_{ivd}\}$.
It is analogous to processing the data blocks on the second and the third diagonals.
Because the global parameter $\Psi_k$ is needed for all block computation,
we store a local copy of $\Psi_k^m$ in each sampling server and synchronize $\Psi_k^m$ by a coordinator server after processing each diagonal of blocks.
We refer to each of non-conflicting $M$ shards as a \emph{segment}.
Figure~\ref{fig:config} shows three segments, $\{1,1,1\}$, $\{2,2,2\}$ and $\{3,3,3\}$, when $M=3$.

The scalability of one configuration in Figure~\ref{fig:config} is limited.
Increasing the number of sampling servers $M$ indicates the increasing of vertical partitions
of the training corpus $\V{x}_{D \times V}$ as well as the model parameters $\Phi_{V \times K}$ in rows.
So,
the number of data servers $M$ would be less than the size of vocabulary $V$.
However,
the size of vocabulary words in one sampling server should be larger than a value (e.g., $\ge 10^3$) for a better efficiency.
In this case,
$M$ cannot be very large in practice.
When $V \ll D$,
it is difficult to use the limited $M$ data servers to store big data in one configuration.
As a result,
we need to build multiple configurations and use a set of $M$ {\em aggregation servers}
to synchronize the global model parameter $\Phi_{vk}$ from different configurations.

Figure~\ref{fig:layer2} shows the hierarchical distributed architecture containing two layers.
In the layer $1$,
there are $1 \le c \le C$ configurations as shown in Figure~\ref{fig:config}.
For simplicity,
we do not illustrate the data servers in each configuration.
In the layer $2$,
there are $M$ aggregation servers to connect corresponding sampling servers.
At the end of each GS iteration,
all sampling servers in all configurations report their model parameter change $\Delta\Phi_{vk}^m$ to aggregation servers~\cite{ad-lda,Ahmed:12}.
Notice that the $m$th sampling server in the layer $1$ configuration reports only to the $m$th aggregation server in the layer $2$.
After the aggregation from all configurations,
the aggregation servers distribute the updated global model $\Phi_{V \times K}$ to all configurations in the layer $1$.
In practice,
the aggregation does not have to be done in every iteration.
Each configuration can run independently for several iterations before the model aggregation in the layer $2$.
This strategy is analogous to the Robbins-Monro stochastic optimization~\cite{Robbins:51},
where a minibatch estimate of model parameters can be viewed as a stochastic approximation to them.
Therefore,
this asynchronous parameter update method in Peacock is likely to work
because Robbins-Monro stochastic optimization works.
Previous results~\cite{ad-lda,Yan:14} also confirm that the
asynchronous and delayed synchronization of the global parameters
will not degrade LDA accuracy very much.

In the layer $2$,
we also use a \emph{coordinator} to control the cooperative work of sampling, data and aggregation servers.
The coordinator is also responsible for updating and re-distributing the global copy of $\Psi_k$ to sampling servers,
and optimizing the asymmetric Dirichlet prior $\alpha_k$.
To optimize $\alpha_k$,
the coordinator has to maintain a histogram of document lengths, $l_d$,
and collects a matrix, $\Omega_{kn}$,
storing the number of documents in which topic $k$ appears $n$ times~\cite{rethinking} from the data servers.
Likewise to the synchronization of $\Phi^{mc}_{vk}$ in different configurations,
we estimate these global parameters in the coordinator and distribute them to sampling servers in the asynchronous manner.
For example,
after all configurations run a few iterations,
the coordinator aggregates $\Psi^{m}_{k}$ (the aggregation cost is small due to a simple sum operation)
and broadcasts the updated global parameter to all sampling servers.
All these servers are developed using Google's Go programming language,
a compiled concurrent system programming language.
The parallel machine learning systems could benefit a lot from the native support of concurrent programming and convenient
implementation of remote procedure calls (RPC) provided by Go.

\subsubsection{Distributed GS Algorithm}

\begin{figure}[t]
\centering
\begin{algorithmic}[1]
\Procedure{RunGibbsIteration}{$M$, $M$}
\For{$\text{segment} \gets 1 .. \lceil M \rceil$} \State \Call{SampleSegment}{segment} \EndFor
\State $\alpha_k \gets $ \Call{OptimizeHyperparameters}
{$\alpha_k$, $l_d$, $\Omega_{kn}$}
\ParFor{$m \gets 1 .. M$} \State sampling-server[$m$].
\Call{SetAlpha}{$\alpha_k$}
\EndProcedure
\Statex \Procedure{SampleSegment}{segment}
\ParFor{$m \gets 1 .. M$}
\State data-server[$m$].
\Call{LoadShard}{segment, $m$} \For{$dig \gets 1 .. M$}
\Comment{$dig$ indices diagonals}
\ParFor{$m \gets 1 .. M$} \\
\qquad\quad data-server[(m+dig)\%M].
\Call{WorkWithSampler}{$m$} as follows:
\For{each token $x_{ivd}^{m}$}
\State{Update $z_{ivd}^{m}$ by sampling one topic from Equation~\eqref{eq:sparseLDA}}
\EndFor
\ParFor{$m \gets 1 .. M$} \State $\Psi_k \gets \Psi_k +$ sampling-server[$m$].
\Call{GetDiffNt}{} \ParFor{$m \gets 1 .. M$}
\State sampling-server[$m$].\Call{SetNt}{$\Psi_k$} \EndFor
\ParFor{$m \gets 1 .. M$} \State
data-server[$m$].\Call{SaveShard}{segment, $m$} \ParFor{$m \gets 1
.. M$} \State $\Omega_{kn} \gets \Omega_{kn} +$
data-server[$m$].\Call{CountNtn}{} \EndProcedure
\end{algorithmic}
\caption{\label{fig:algo} The distributed GS algorithm.}
\end{figure}

Figure~\ref{fig:algo} shows the distributed GS algorithm executed by the coordinator,
where the dot symbol denotes an RPC call.
For example,
in line~$6$,
the coordinator calls procedure \textsc{SetAlpha},
which is exposed and executed by a sampling server,
where the sampling follows Equation~\eqref{eq:sparseLDA}.
The \textbf{par-for} in Figure~\ref{fig:algo} denotes the concurrent version of the commonly-used control structure \textbf{for}.
The \textbf{par-for} flattens the loop body and executes it in parallel.
The \textbf{par-for} can be implemented using Go language elements of \emph{channel} and \emph{goroutine},
as shown by the open source project \url{https://github.com/wangkuiyi/parallel},
which is used in Peacock.
C/C++ programmers can use the \textbf{par-for} provided by OpenMP.

In each iteration,
the procedure \textsc{RunGibbsIteration} invokes \textsc{SampleSegment} to update the topic assignments segment by segment,
where \textsc{SampleSegment} coordinates the $M$ data servers and the $M$ sampling servers to run the parallel SparseLDA algorithm.
Each sampling server keeps a local $\Phi^m_{vk}$ and a global $\Psi_{k}$,
and updates the topic assignment $z^m_{ivd}$ sent by the data server in line $14$.
\textsc{SampleSegment} also collects the matrix,
$\Omega_{kn}$,
which records the number of documents in which the topic assignment $k$ occurs for $n$ times.
This matrix is used by the procedure \textsc{OptimizeHyperparams},
which is described in~\cite{rethinking},
to optimize the asymmetric prior, $\alpha_k$,
at the end of each GS iteration.
In addition to $\Omega_{kn}$,
\textsc{OptimizeHyperparameters} also requires $l_d$,
the vector recording the document lengths,
which is counted in the first iteration and saved in coordinator for later use.

\subsubsection{Pipeline for Efficient Communication} \label{communication}

\begin{table}[t]
\tbl{The parameters of the communication pipeline and the corresponding communication time.}
{\begin{tabular}{r|r|r}
\hline
$T$ & $L$ (KB) & Time (minutes) \\
\hline
    200,000 &       1 & 48.1 \\
     20,000 &      10 & 45.3 \\
      2,000 &     100 & 43.5 \\
        200 &   1,000 & 43.3 \\
         40 &   5,000 & 43.4 \\
         20 &  10,000 & 43.5 \\
         10 &  20,000 & 44.1 \\
          1 & 200,000 & 49.8 \\
\end{tabular}}
\label{tab:pipeline}
\end{table}

Figure~\ref{fig:algo} shows that all network communications in Peacock happen in the form of RPCs.
The largest fraction of {\em communication cost} lies in \textsc{WorkWithSampler},
where the data servers send data blocks to the sampling servers,
and wait for responses containing the updated topic assignment $\V{z}^m_{ivd}$.
We reduce the communication cost using the \emph{pipeline} technique.
To avoid the overflow of network communication buffer,
the data server sends just a few document fragments known as a \emph{package}
rather than sending a block in an RPC to the sampling server.
Instead of waiting for the response from the sampling server before sending the next package,
the data server sends $T$ packages concurrently.
On the sampling server, there are multiple \emph{goroutines},
a kind of light-weighted thread scheduled by the Go runtime system,
to process these packages and respond to the data server.
The data server maintains a data structure with $T$ slots,
and each keeps track of an out-going package.
The data server clears a slot after receiving a response of the corresponding package,
or getting a timeout.
Once there are empty slots and packages to be processed,
the data server would continue sending packages.
In general,
this pipeline optimizes the throughput by overlaps the sending, processing and responding of packages.

Maximizing the throughput depends on finding the optimal configuration of two parameters:
package size $L$ and pipeline capacity $T$.
The product,
$L \times T$,
is proportional to the size of memory used as communication buffer.
In practice,
there would be an upper limit of buffer size,
$y$,
and we would make full use of it to get the maximum throughput.
This can be written as the constraint function,
$y = L \times T$,
which is a curve on the two dimensional space $L$ and $T$.
The best configuration would be a point on this curve.
In our computing environment,
a practical $y$ value is $200$MB.
The measure of time consumption with respect to the curve $y = L \times T$ is shown in Table~\ref{tab:pipeline},
where the time consumption is larger at both ends of this curve,
and the optimal configuration lies in the middle of the curve.
The variance is too small to be shown.

\subsubsection{Lock-free Synchronization} \label{synchronization}

\begin{figure}[t]
\begin{algorithmic}[1]
\Procedure{SampleSegment}{segment} \ParFor{$m \gets 1 .. M$} \State
data-server[$m$].\Call{LoadShard}{segment, $m$} \ParFor{$m \gets 1
.. M$} \For{$dig \gets 1 .. M$}  \Comment{$dig$ indices diagonals}
\State data-server[(m+dig)\%M].\Call{WorkWithSampler}{$m$} \EndFor
\ParFor{$m \gets 1 .. M$} \State $\Psi_k \gets \Psi_k +$
sampling-server[$m$].\Call{GetDiffNt}{} \ParFor{$m \gets 1 .. M$}
\State sampling-server[$m$].\Call{SetNt}{$\Psi_k$} \ParFor{$m \gets
1 .. M$} \State data-server[$m$].\Call{SaveShard}{segment, $m$}
\ParFor{$m \gets 1 .. M$} \State $\Omega_{kn} \gets \Omega_{kn} +$
data-server[$m$].\Call{CountNtn}{} \EndProcedure
\end{algorithmic}
\caption{The Faster sampling of corpus segments.}
\label{fig:per-segment}
\end{figure}

After issuing parallel executions of the loop body in Figure~\ref{fig:algo},
the \textbf{par-for} does a synchronization operation that waits for
the completions of all executions of sampling servers called the {\em synchronization lock problem}.
We address this problem by three lock-free strategies.
First,
to avoid data skewness in each data block,
we randomly shuffle data $\V{x}_{D \times V}$ by rows and columns so that each block contains almost equal number of word tokens in practice~\cite{Zhuang:13}.

Second,
we can further reduce the synchronization cost of the \textbf{par-for}
on line~$4$ of Figure~\ref{fig:per-segment} by balancing the workload of sampling servers.
Because each sampling server processes a row of corpus blocks,
it is desirable that the block rows contain similar word frequencies.
This can be achieved using a pre-training scheduler,
which assigns vocabulary words to $\Phi_{vk}^m$ shards.
As with PLDA+~\cite{plda+},
we use the weighted round-robin method for this word assignment.
We first sort vocabulary words in descending order by their frequency,
and pick the word with the largest frequency and assign it to the $\Phi_{vk}^{m}$ shard with the accumulative word frequency.
Then,
we update the accumulated word frequency of $\Phi_{vk}^{m}$.
This placement process is repeated until all words have been assigned.
Weighted round-robin has been empirically shown to achieve a balanced load with a high probability~\cite{load-balancing}.

Finally,
in Figure~\ref{fig:algo},
the nested loop starting from line~$10$ invokes the \textbf{par-for} many times and introduces many waits.
The problem can be relieved by swapping the inner and outer loop as shown in Figure~\ref{fig:per-segment}.
In this change,
two \textbf{par-for} structures at line~$13$ and line~$15$ are moved one upper level,
which relaxes the aggregation and redistribution of vector $\Psi_k$ from a per-diagonal granularity to a per-segment granularity.
This relaxed aggregation does not affect the correctness of the distributed GS algorithm
analogous to the stochastic optimization framework~\cite{Robbins:51}.
Indeed,
similar asynchronous optimization methods for updating model parameters
have been confirmed to work well in lock-free parallel stochastic gradient descent algorithms~\cite{Niu:11,Johnson:13}.
After swapping the \textbf{par-for} at line~$11$ with its outer loop at line~$10$,
a data server might work with more than one sampling server simultaneously.
However,
this would introduce conflicts in accessing the $\V{z}_{ivd}^m$ shard maintained by the data server.
We address this conflict problem by two methods.
First,
we use $M+1$ data servers as shown in Figure~\ref{fig:config} denoted by ``Free".
These ``Free" data servers provide additional conflict-free data blocks for computing without waiting for the completion of other sampling servers.
The coordinator will schedule the finished sampling server to those conflict-free data blocks having the minimum number of visits.
In this way,
we assure that all data blocks will have almost equal number of visits.
Second,
each data server maintains two copies of the $\V{z}_{ivd}$ shard:
$\V{z}_{ivd}^{old}$ and $\V{z}_{ivd}^{new}$ without conflicts.
After receiving the response of updated package from the sampling server,
the data server applies the difference between the response and $\V{z}_{ivd}^{old}$ to $\V{z}_{ivd}^{new}$.

\subsubsection{Fault Recovery}

Data parallelism also helps fault recovery,
which is critical in large-scale machine learning.
Consider that a parallel learning job may take days or even weeks,
it is very likely that some workers fail or be preempted during the period.
If the system cannot recover when it fails,
we would have to restart the job from the beginning.
Since the restart might fail again,
the learning job would never finish.
As the configurations in the layer $1$ work independently within every few iterations,
it is straightforward to restart any failed configuration based on the independent checkpoint on hard disks.
The restart can be implemented simply using the SSH command,
or sophisticated cluster management systems like Apache YARN.
This architecture is similar to Google's architecture for parallel deep learning~\cite{ddnn},
which refers to configurations as models.
More than achieving fine-grained fault recovery,
this design also improves the parallelism and makes Peacock highly scalable.

\subsection{Real-time Prediction} \label{sec:real-time-inference}

\begin{figure}
\centerline{
\includegraphics[width=0.9\linewidth]{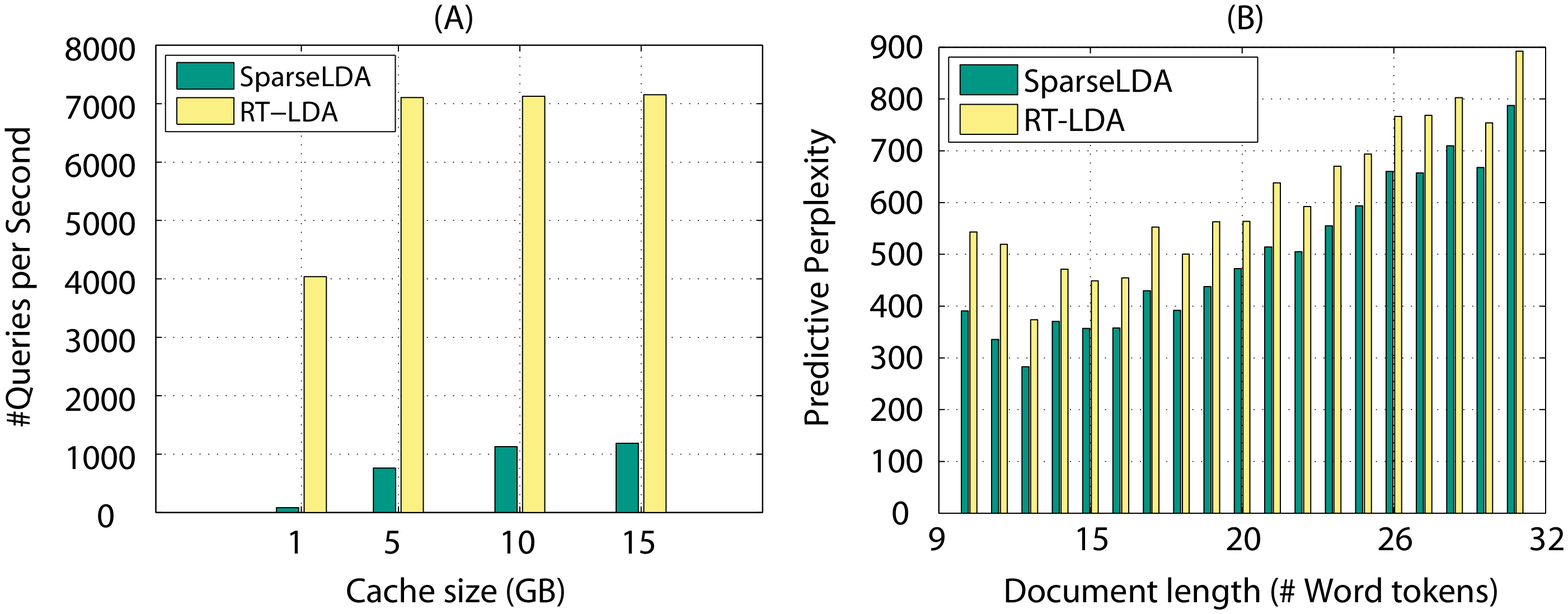}}
\caption{Comparisons between RT-LDA and SparseLDA~\cite{sparselda}
in (A) speed and (B) accuracy.} \label{fig:rt-lda}
\end{figure}

It is critical in online applications like search engine and online advertising systems to predict latent semantics of new user queries
in real-time based on the large number of topics.
In typical Internet services,
the response time of backend servers is measured in milliseconds.
Given LDA models with at least $10^5$ topics,
few inference algorithms are efficient enough to do real-time prediction.
Hence,
we propose a real-time inference algorithm,
RT-LDA,
especially for prediction of new queries.

The basic idea of RT-LDA is to replace the sampling operation in
Equation~\eqref{eq:sparseLDA} by the $\max$ operation.
This makes RT-LDA a hill climbing or coordinate ascent algorithm whose search path consists of line
segments aligned with axes of the topic space,
which is similar to the coordinate ascent using one-dimensional Newton step~\cite{cdn} widely used in learning regression models.
The $\max$ operation in RT-LDA can be optimized using a cache-based technique.
According to Equation~\eqref{eq:gibbs-sampling},
the $\max$ operator in RT-LDA is
\begin{equation}
\label{eq:hill-climb}
\begin{split}
& \max_{k\in[1,K]} P(z_{ivd}=k, x_{ivd}=1 \mid \V{z}_{\neg ivd}, \V{x}_{\neg ivd}, \alpha, \beta) \\
= & \max_{k\in[1,K]} \phi_{vk} (\Theta_{dk} + \alpha_k)  \\
= & \max_{k\in[1,K]} \phi_{vk} \Theta_{dk} + \phi_{vk} \alpha_k,
\end{split}
\end{equation}
where $\phi_{vk}$ is the empirical probability matrix computed by Equation~\eqref{phi}.
In prediction,
$\phi_{vk}$ and $\alpha_k$ are constants,
whereas $\Theta_{dk}$ changes with the updating of topic assignments.
This makes it viable to precompute $\max_k{\phi_{vk}\alpha_k}$,
whose result is a sparse matrix $\M{R}$,
\begin{equation} \label{eq:2}
R_{vk} = \begin{cases}
\phi_{vk}\alpha_{k} & \text{if } k=\max_{k'}{\phi_{vk'}\alpha_{k'}}, \\
0 & \text{otherwise}.
\end{cases}
\end{equation}
We save $\M{R}$ in a compact data structure which contains only $V$ non-zero elements.
The compact $\M{R}$ is an approximation to Equation~\eqref{eq:hill-climb},
where the error of the approximation is caused by non-zero elements in $\Theta_{dk}$.
We rewrite Equation~\eqref{eq:hill-climb} as
\begin{equation} \label{eq:rt-lda}
\begin{split}
& \max_{k\in[1,K]} P(z_{ivd}=k, x_{ivd}=1 \mid \V{z}_{\neg ivd}, \V{x}_{\neg ivd}, \alpha, \beta) \\
= & \max_k \left[ R_{vk}^*, \max_{\substack{k\in[1,K] \\ s.t.
\Theta_{dk}>0}} \phi_{vk} (\Theta_{dk}+ \alpha_k) \right],
\end{split}
\end{equation}
where $R_{vk}^*$ denotes the non-zero element in the column of $\M{R}$ corresponding to the vocabulary word $v$.
Different from the $\max$ operation in Equation~\eqref{eq:hill-climb},
which iterates over $k\in[1,K]$,
the first $\max$ operation in Equation~\eqref{eq:rt-lda} compares two values,
and the second $\max$ operation visits only non-zero elements in $\Theta_{dk}$.
Suppose that the maximum number of non-zero elements in a query $d$ is the length of the query,
Equation~\eqref{eq:rt-lda} makes RT-LDA significantly faster than SparseLDA when the number of topics $K \ge 10^5$.

Figure~\ref{fig:rt-lda}A compares the prediction speed between RT-LDA and SparseLDA.
We use the the number of query-per-second (QPS) of RT-LDA and SparseLDA on a real backend inference server with $100\%$ CPU load.
The $x$-axis is the cache size.
Generally,
the larger cache leads to faster prediction,
but the performance reaches the upper bound with the increase of cache size in Gigabytes (GB).
This setting implies that the response time of prediction is the reciprocal of the QPS.
We see that RT-LDA is about an order of magnitude faster than SparseLDA.
Figure~\ref{fig:rt-lda}B compares RT-LDA and SparseLDA for their topic modeling accuracy measured in the predictive perplexity~\cite{Zeng:11},
which is a standard performance measure for topic modeling accuracy.
The lower perplexity means a higher topic modeling accuracy on new test data set.
This experiment uses $1,200$ Wikipedia titles by clustering them into $20$ groups with various lengths.
For all these groups,
we randomly select $10\%$ as test data set and retain the remaining $90\%$ as training set.
We see that the accuracy of the two algorithms, measured in perplexity, are very close.
RT-LDA loses some tolerable topic modeling accuracy to get a faster speed compared with SparseLDA.
We can further improve the effectiveness of RT-LDA by running multiple line searches in parallel and then averaging results of all these parallel trails.
Because RT-LDA is much more efficient than SparseLDA,
the Peacock system can afford many parallel trails to extract topic features from new queries.

\subsection{Topic De-Duplication} \label{sec:topic-dups}

Although topic duplication has been rarely discussed in previous literature,
it becomes a main challenge in topic feature engineering.
As noted in~\cite{rethinking},
when learning LDA,
frequent words often dominate more than one topics,
and the learned topics are similar to each other.
We refer to these similar topics as \emph{duplicates}.
Generally,
when learning $\ge 10^5$ topics,
around $20\% \sim 40\%$ topics have duplicates in practice.
If a query $d$ is $60\%$ about the topic $A$ and $40\%$ about the topic $B$,
the query $d$ is mainly about the topic $A$.
However,
if a topic $A$ has three duplicates,
$A1$, $A2$ and $A3$,
the GS algorithm would follow Equation~\eqref{eq:gibbs-sampling} to scatter the $60\%$ weight of $A$ in $d$ to $A1$, $A2$ and $A3$.
If each duplicate gets $1/3$ of the original $60\%$ topic $A$,
the interpretation of the query would become mainly about topic $B$,
though the truth is that the query is mainly about topic $A$.

We remove topic duplicates by two methods.
First,
we learn the asymmetric Dirichlet prior $\alpha_k$ over the document-topic distributions~\cite{rethinking},
which substantially increases the robustness of LDA to variations in the number of topics and to the
highly skewed word frequency distributions common in natural language.
Asymmetric priors over document-topic distributions automatically combine similar topics into one large topic,
rather than splitting topics more uniformly by symmetric priors.
In practice,
we can set a very large $K=10^6$ value at initial learning time,
and prune duplicates by the asymmetric Dirichlet prior.
Those topics with very small Dirichlet priors would be automatically weighted trivial by RT-LDA (Subsection~\ref{sec:real-time-inference}) at serving time.
We find this approach prevents common words from dominating many topics,
thus leaves the room for long-tail topics.
Second,
we cluster topic duplicates if their $L_1$-distance is below a threshold.
The lower $L_1$-distance threshold means that we would remove more duplicates from large number of topics.

\section{Empirical Studies in Big Data} \label{sec:parallel}

We evaluate Peacock's topic modeling performance for big data by three performance measures:
1) Speedup: the runtime ratio over a sequential GS algorithm as we increase the number of computing cores available;
2) Scalability: the ability to handle a growing number of topics;
3) Accuracy: the held-out log-likelihood of LDA achieved by increasing number of iterations~\cite{smola,Ahmed:12}.
The baseline is the state-of-the-art industrial solution
Yahoo!LDA~\cite{smola,Ahmed:12} with open source codes\footnote{\scriptsize
\url{https://github.com/sudar/Yahoo_LDA}}.
Likewise,
Yahoo!LDA also distributes SparseLDA~\cite{sparselda} over multiple machines but using a shared memory environment based on the \emph{memcached} technique.
In the layer $2$ of Peacock,
we use $1$ coordinator server and $M = 125$ aggregation servers (as shown in Figure~\ref{fig:layer2}).
In the layer $1$,
we set $C=2$ configurations,
each of which contains $M = 125$ sampling servers and $M + 1 = 126$ data servers.

\subsection{Data Sets} \label{sec:data}

For a fair comparison, we use the same publicly available data set PUBMED\footnote{\scriptsize
\url{http://archive.ics.uci.edu/ml/datasets/Bag+of+Words}},
which contains $8.2$ million documents with an average length of around $90$ word tokens each.
The vocabulary size of PUBMED is $1.4 \times 10^5$.
We also compose the training corpus of search queries received in recent months called SOSO.
The pre-processing of the corpus contains five steps:
1) Transform each query into word tokens, and count word frequencies.
2) Remove those words with low frequency, which are likely typos.
3) Remove those words with very high frequency,
because common words tend to dominate all topics~\cite{rethinking}.
4) De-duplicate queries:
If a query appears multiple times,
we keep only one appearance in corpus.
This allows us to include a large variety of user intentions within a certain amount of training corpus.
This also lowers the weight of frequent queries in the corpus.
5) Remove those queries containing only one word,
because single-word queries do not provide word co-occurrence counts,
which is a clue used by LDA to infer topics.
The processed corpus contains one billion search queries with $4.5$ word tokens per query on average and takes $17.2$GB storage space.
The vocabulary size of SOSO is around $2.1 \times 10^5$.
Obviously,
SOSO is around $6$ times larger than PUBMED.

\subsection{Results}

\begin{figure}[t]
\centerline{
\includegraphics[width=0.8\linewidth]{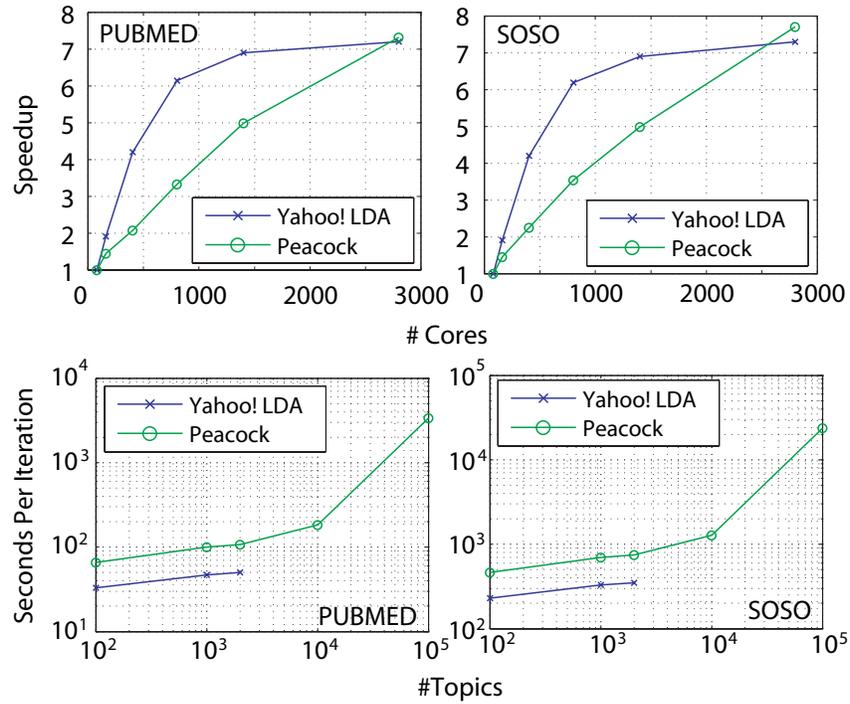}}
\caption{\label{fig:speed}
The speedup and scalability performance.}
\end{figure}

\begin{figure}[t]
\centerline{
\includegraphics[width=0.8\linewidth]{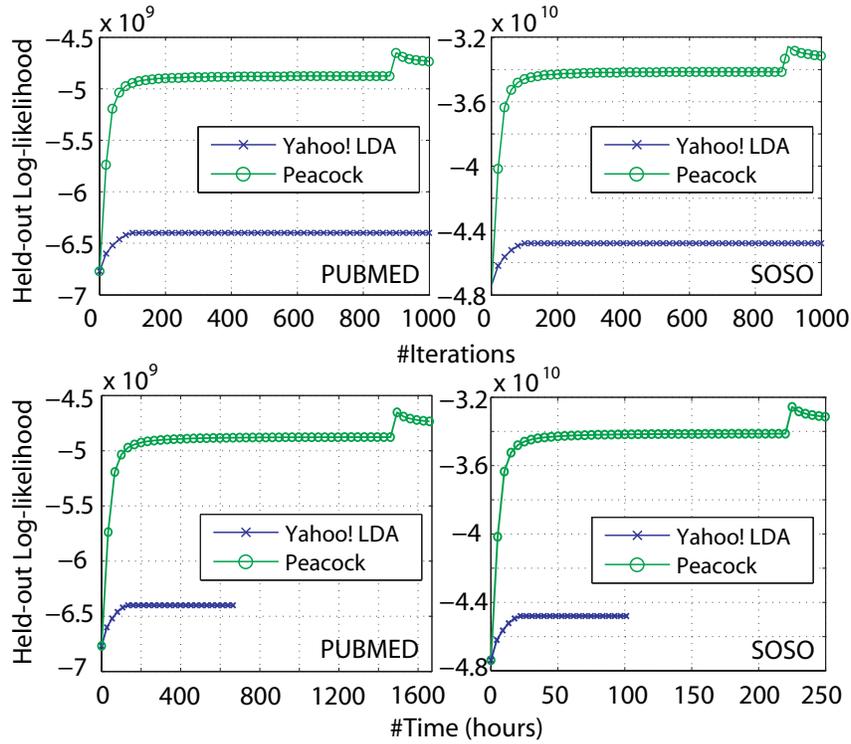}}
\caption{\label{fig:logllhood}
The topic modeling accuracy and convergence speed.}
\end{figure}

\begin{figure}[t]
\centerline{
\includegraphics[width=0.8\linewidth]{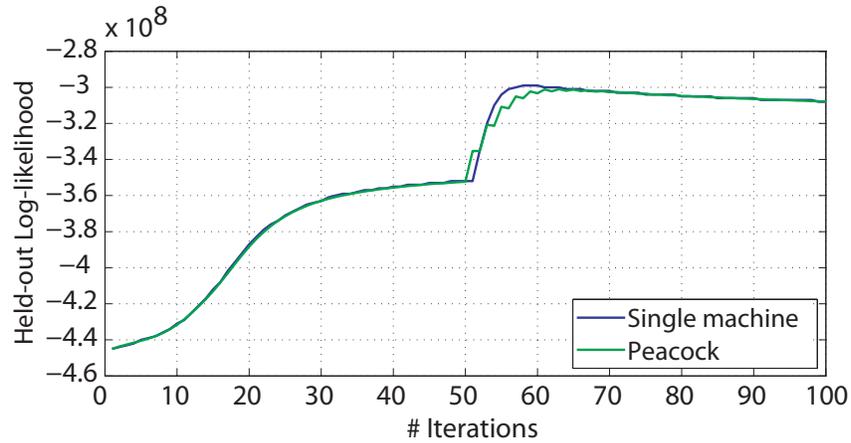}}
\caption{\label{fig:single}
Comparisons between Peacock and the single machine.}
\end{figure}

Figure~\ref{fig:speed} shows the speedup performance
(fixing $K=1000$),
where the $x$-axis is the number of cores and $y$-axis is the runtime ratio of $100$ cores over that of other number of cores in $x$-axis.
The variance is too small to be shown.
We see that Peacock on average achieves around $4.2$ speedup when the number of cores is $1000$,
which implies that the communication and synchronization in Peacock take about half of the training time.
Yahoo!LDA has a much better speedup than Peacock when the number of cores is small ($\le 1000$).
However,
its speedup performance drops significantly when the number of cores increases from $1000$ to $3000$.
The possible reason is that Yahoo!LDA does not consider the lock-free synchronization problem.
In practice,
the very large number of cores will cause longer waiting time when accessing the shared memory based on the \emph{memcached} technique.
Peacock scales much better to the large number of cores by pipeline communication and lock-free synchronization
(Subsections~\ref{communication} and~\ref{synchronization}).
We see that Peacock is slower than Yahoo!LDA when the number of cores is small.
The reason is that we use more separate sampling servers for model parallelism leading to the additional communication and synchronization costs,
which remains almost a constant ratio in training time by pipeline techniques.

Figure~\ref{fig:speed} also shows the training time per iteration with the increasing number of topics
(fixing the number of cores to $500$),
$K = \{10^2, 10^3, 10^4, 10^5\}$.
With $K$ increasing from $10^2$ to $10^4$,
the corresponding training time increase of Peacock is small.
When $K$ increases by $10$ times from $10^4$ to $10^5$,
the training time is close to the linear growth.
The scalability of Peacock with respect to $K$ comes mainly from the model and data parallelism (Subsection~\ref{sec:hierarchical})
resulting in very fast sampling performance with a small memory footprint.
From $K = 10^4$ to $K = 10^5$,
Peacock consumes significantly more training time because the topic sparseness of each word token becomes lower in SparseLDA.
As a comparison,
Yahoo!LDA has out of memory problem when the the number of topics $K \ge 10^4$ because it does not consider storing a big topic-word count matrix $\Phi_{vk}$.
Peacock divides this big parameter matrix into small model shards to handle increasing number of topics.
Since Peacock uses additional communication to synchronize more sampling servers,
its per iteration training time is slightly longer than that of Yahoo!LDA.

We use the iteration annealed importance sampling (iteration-AIS)~\cite{Wallach:09,Foulds:14} method
to evaluate the predictive performance (measured by the held-out log-likelihood) of Peacock.
We randomly select $10^4$ documents from PUBMED and $10^5$ queries from SOSO as the held-out data sets.
Figure~\ref{fig:logllhood} shows the held-out log-likelihood as a function of training iterations and time
(fixing $K=1000$ and the number of cores $500$) while variance is too small to be shown.
The higher held-out log-likelihood corresponds to the better model quality.
We observe that Peacock has a rise after $900$ iterations
because we start asymmetric prior optimization and topic de-duplication (Subsection~\ref{sec:topic-dups}),
which can improve the topic model quality.
Although both Peacock and Yahoo!LDA use SparseLDA,
Peacock converges to a higher log-likelihood level.
The reason is partly because Yahoo!LDA uses the approximate synchronization to speedup its performance leading to a slightly worse model quality,
which has been also observed in their own work~\cite{smola,Ahmed:12}.
Figure~\ref{fig:logllhood} also shows that Peacock uses less training time to achieve a higher held-out log-likelihood than Yahoo!LDA.
To see if Peacock can produce the same model quality as that generated by a sequential GS inference on a single machine,
we show their held-out log-likelihoods as a function of training iterations in Figure~\ref{fig:single} while variance is too small to be shown.
Since the single machine cannot train the large number of samples due to the memory constraint,
we randomly select a subset of the training set $10^6$ queries,
and select $10^4$ queries as the held-out set.
We see that the held-out log-likelihood curve generated by Peacock locates closely to that produced by the sequential GS on the single machine.
This result confirms that the approximate synchronization techniques used in Peacock do not affect the model quality very much.

To summarize,
we see that Yahoo!LDA is more efficient for small-scale ($K \approx 10^3$) topic modeling tasks,
while Peacock is more suitable for solving large-scale ($K \ge 10^5$) topic modeling problems in industrial applications.

\section{Online Applications} \label{sec:industry}

After Peacock learns $K \ge 10^5$ topics from $D \ge 10^9$ queries,
we need to integrate the topic features into existing search engine
and online advertising systems. We extract topic features from new
queries based on the topic distribution over words $\phi_{vk}$ in Equation~\eqref{phi} learned by Peacock.
Given the word tokens of a new query $d$, we use RT-LDA to predict
its topic distribution $\theta_{dk}$ by fixing $\phi_{vk}$. Employing the
Bayes' rule, we calculate the likelihood $P(v|d)$ of a vocabulary
word $v$ given a query $d$:
\begin{align}\label{eq:pwd}
P(v|d)=\sum_{k=1}^{K} \phi_{vk}\theta_{dk}.
\end{align}
The $V$-length vector $P(v|d)$ is compatible with the standard word vector space model.
If we rank $P(v|d)$ in descending order,
we obtain top likely topic features in the input query.

\subsection{Experimental Settings}

Search engine uses the well-known vector space model in information retrieval and
compute cosine similarity between queries and documents in their vector representations.
We accelerate this process by using the Weak-AND algorithm~\cite{wand}.
Peacock replaces the word vector features of each query by top $30$ likely topic features in Equation~\eqref{eq:pwd}
(Top $30$ largest values from $V$-length vector $P(v|d)$)
inferred by RT-LDA in the head of each posting list used by the Weak-AND algorithm,
which makes the query-document similarity computing efficient enough to be deployed in a real search engine.

Online advertising has been a fundamental financial support of the
many free Internet services~\cite{ad-lecture}.
Most contemporary online advertising systems follow the Generalized Second Price (GSP) auction model~\cite{gsp},
which requires that the system is able to predict the click-through rate (pCTR) of an ad,
where pCTR is an important clue in GSP to ranking ads and pricing clicks.
One of the key questions with the pCTR is the availability of suitable input features or
predictor variables that allow accurate CTR prediction for a given ad impression~\cite{features}.
These features can be grouped into three categories:
ad features including bid phrases, ad title, landing page, and a hierarchy of advertiser account, campaign, ad group and ad creative.
User features include recent search queries,
and user behavior data.
Context features include display location,
geographic location,
content of page under browsing,
and time.
Most of these features are text data in word vector space~\cite{ad-predictor}.
We learn an $L_1$-regularized log-linear model~\cite{owl-qn} as the baseline for pCTR,
which uses a set of text and other features such as ad title,
content of page under browsing, content of landing page,
ad group id,
demographic information of users,
categories of ad group and categories of the page under browsing.
As a comparison,
Peacock adds all topic features~\eqref{eq:pwd},
i.e.,
the $V$-length topic feature vector $P(v|d)$,
in baseline text and other features as input to $L_1$-regularized log-linear model~\cite{owl-qn}.

In both applications,
the training data set of Peacock is SOSO described in Subsection~\ref{sec:data}.
For Peacock,
we set $C=2$ configurations with $1$ coordinator server and $M = 125$ aggregation servers.
Each configuration contains $M = 125$ sampling servers and $M + 1 = 126$ data servers.

\subsection{Results}

\begin{figure}[t]
\centerline{
\includegraphics[width=0.7\linewidth]{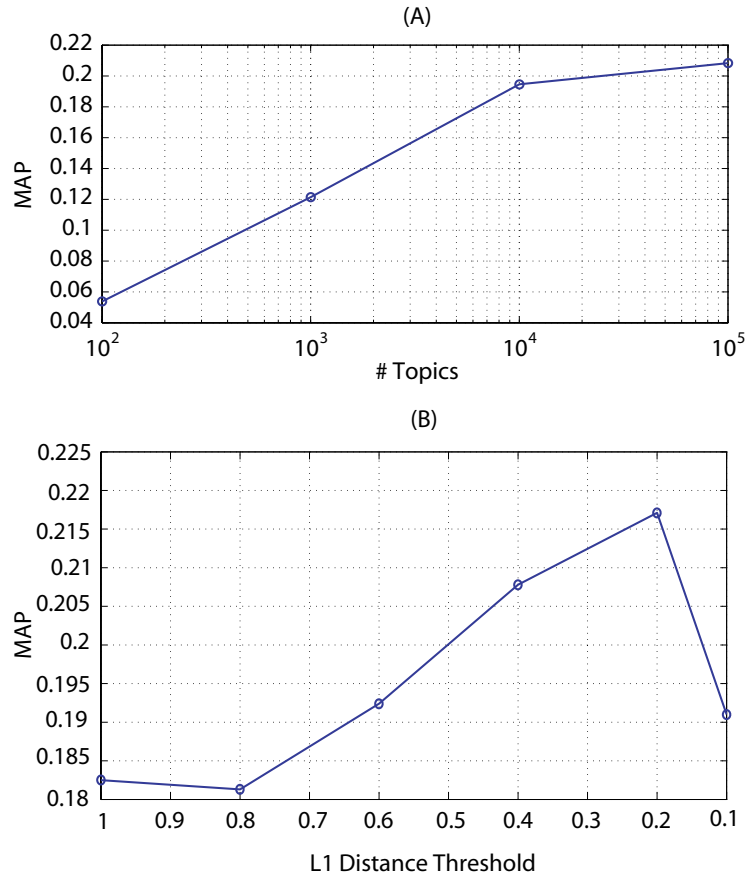}}
\caption{(A) Topic features improve retrieval in search engine.
(B) Performance improvement in retrieval after topic de-duplication by topic clustering based on $L_1$ distance.}
\label{fig:retrieval}
\end{figure}

For information retrieval,
our test-bed is a real search engine,
\url{www.soso.com},
which ranks the fourth largest in China market.
The test data is used for routinely relevance evaluation,
containing $4,818$ randomly selected queries and $121,588$ query-URL pairs with human labeled relevance rate.
Every query-URL pair was rated by three human editors and the average rate was taken.
We compute mean average precision (MAP) using TREC evaluation tool~\cite{trec-book}.
The higher MAP means the better retrieval performance.
Figure~\ref{fig:retrieval}A shows that the topic features improve the relevance measure by MAP with small variance.
The relevance improvement grows steadily with the increasing number of topics from $10^2$ to $10^5$.
However,
the growth of MAP becomes less salient when the number of topics changes from $10^4$ to $10^5$.
This is mainly attributed to the problem of \emph{topic duplication}.
Figure~\ref{fig:retrieval}B shows that topic de-duplication method can further improve the relevance of information retrieval.
The MAP value of retrieval grows when we prune more duplicated topics (lower $L_1$ distance can prune more similar topics in Subsection~\ref{sec:topic-dups}).
Usually,
the MAP stops increasing when we prune duplicates from the initial $10^6$ to around $10^5$ topics.
This result implies that $10^5$ is a critical number of topics to describe subtle word senses in big query data with $2.1 \times 10^5$ vocabulary words.
If the $L_1$ distance is too small such as $0.1$,
it will degrade the MAP performance by removing more non-duplicate topics.

\begin{figure}
\centerline{
\includegraphics[width=0.7\linewidth]{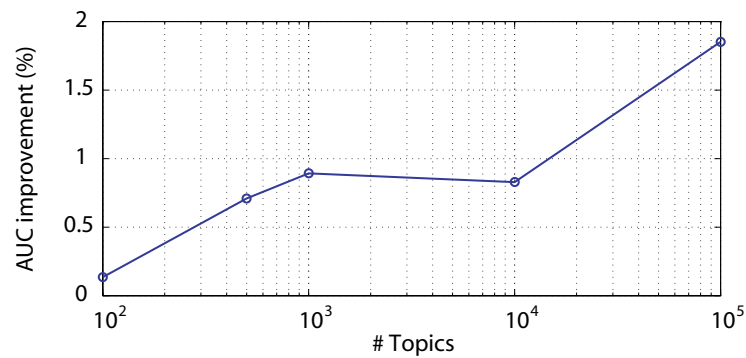}}
\caption{Topic features improve the pCTR performance in online advertising systems.}
\label{fig:advertising}
\end{figure}

The online advertising experiment is conducted on a real contextual advertising system,
\url{https://tg.qq.com/}.
This system logs every ad shown to a particular user in a particular page view as an \emph{ad impression}.
It also logs every click of an ad.
By taking each impression as a training instance,
and labeling it by whether it was clicked,
we obtain $9.9$ billion training samples and $1.1$ billion test samples.
We train $5$ \emph{hypothetical models},
whose topic features are extracted using $5$ different LDA models with $10^2$, $5 \times 10^2$, $10^3$, $10^4$ and $10^5$ topics,
respectively.
Following the judgment rule of Task $2$ in KDD Cup $2012$,
a competition of ad pCTR, we compare our hypothetical models with the baseline by their prediction performance measured in area under the curve (AUC).
Figure~\ref{fig:advertising} shows that all hypothetical models gain relative AUC improvement ($\%$) than the baseline (AUC $= 0.7439$).
Variance is too small to be shown.
This verifies the value of big LDA models.
Also,
the AUC improvement grows with the increase of the number of topics learned by Peacock.
The reason that the performance of $10^4$ is lower than that of $10^3$ is because of many topic duplicates in $10^4$ topics.
After using automatic topic de-duplication by asymmetric Dirchlet prior learning (Subsection~\ref{sec:topic-dups}),
the performance of $10^5$ becomes better than that of $10^4$.
This result is consistent with those in Figure~\ref{fig:retrieval}.

\section{Conclusions} \label{sec:conclusion}

Topic modeling techniques for big data are needed in many real-world applications.
In this paper,
we confirm that a big LDA model with at least $10^5$ topics inferred from $10^9$ search queries can achieve
a significant improvement in industrial applications like search engine and online advertising systems.
We propose a unified solution Peacock to do topic modeling for big data.
Peacock uses a hierarchical distributed architecture to handle large-scale data as well as LDA parameters.
In addition,
Peacock addresses some novel problems in big topic modeling,
including real-time prediction and topic de-duplication.
We show that Peacock is scalable to more topics than the current state-of-the-art industrial solution Yahoo!LDA.
Through two online applications,
we also obtain the following experiences:
\begin{itemize}
\item
The good performance is often achieved when the number of topics is approximately equal to or more than the number of vocabulary words.
In our experiments,
the vocabulary size is $2.1 \times 10^5$ so that the number of topics $K \ge 10^5$.
In other industrial applications,
the vocabulary size may reach a few millions or even a billion.
The Peacock system can do topic feature learning when $K \ge 10^7$ is needed.
\item
The topic de-duplication method is a key technical component to ensure that $K \ge 10^5$ topics can provide high-quality topic features.
Better topic de-duplication techniques remain to be an open research issue.
\item
The real-time topic prediction method for a large number of topics is also important in industrial applications.
If $K \ge 10^7$,
faster prediction methods are needed and remain to be a future research issue.
\end{itemize}
In our future work,
we will study how to deploy online LDA algorithms in Peacock,
and how to implement inference algorithms to learn other topic models such as HDP~\cite{hdp} and author-topic models~\cite{Steyvers:04}.

\begin{acks}
This work was supported by National Grant Fundamental Research (973 Program) of China under Grant 2014CB340304,
NSFC (Grant No. 61373092 and 61033013),
Natural Science Foundation of the Jiangsu Higher Education Institutions of China (Grant No. 12KJA520004),
and Innovative Research Team in Soochow University (Grant No. SDT2012B02).
This work was partially supported by Collaborative Innovation Center of Novel Software Technology and Industrialization.
\end{acks}





\medskip

\end{document}